\documentclass[twocolumn,showpacs,preprintnumbers]{revtex4}%
\usepackage{amssymb}
\usepackage{amsmath}
\usepackage{graphicx}
\usepackage{dcolumn}
\usepackage{bm}
\usepackage{amsfonts}%
\begin{document}
\title{Decoherence and Recoherence in a Vibrating RF SQUID}
\author{Eyal Buks}
\affiliation{Department of Electrical Engineering, Technion, Haifa 32000 Israel}
\author{M. P. Blencowe}
\affiliation{Department of Physics and Astronomy, Dartmouth College, Hanover, New Hampshire
03755, USA}
\date{\today }

\begin{abstract}
We study an RF SQUID, in which a section of the loop is a freely suspended
beam that is allowed to oscillate mechanically. The coupling between the RF
SQUID and the mechanical resonator originates from the dependence of the total
magnetic flux threading the loop on the displacement of the resonator. Motion
of the latter affects the visibility of Rabi oscillations between the two
lowest energy states of the RF SQUID. We address the feasibility of
experimental observation of decoherence and recoherence, namely decay and rise
of the visibility, in such a system.

\end{abstract}
\maketitle



%It is always \today, today,
%but any date may be explicitly specified

%PACS, the Physics and Astronomy
%Classification Scheme.

\section{Introduction}

Decoherence occurs when a quantum system is coupled to a noisy environment at
a finite temperature. Decoherence is commonly quantified by a visibility
factor, which characterizes the relative amplitude of a measured interference
signal. In many cases the main contribution to decoherence originates from the
many degrees of freedom of the environment, which all have a similar coupling
strength to the interfering degree of freedom of the quantum system. In such a
case the visibility factor is expected to decay monotonically as a function of
time (typically, the decay is exponential). On the other hand, when only a few
degrees of freedom in the environment significantly contribute, the time
dependence of the visibility factor is not necessarily monotonic. Recoherence
occurs when the visibility factor increases with time. Experimental
demonstration of this phenomenon is important since it may provide a crucial
test to the theory of quantum measurement \cite{Legget_R415,Leggett_857}.
Decoherence and recoherence were recently discussed theoretically in Refs.
\cite{Armour_035311,Bose_4175,Mancini_3042,Marshall_130401,Bernad_0604157,Armour_148301,Cleland_070501}%
. The interfering quantum system in Ref. \cite{Armour_035311} was a single
level quantum dot, in Refs.
\cite{Bose_4175,Mancini_3042,Marshall_130401,Bernad_0604157} it was an optical
mode in a cavity, and in Refs. \cite{Armour_148301,Cleland_070501} a
superconducting charge (Cooper-pair box) and phase Josephson qubit,
respectively. In all these cases, the interfering quantum system is coupled to
a vibrating mode of a mechanical resonator (typically the lowest, fundamental
mode). Recoherence can occur in such systems provided that the coupling
between the interfering quantum system and the mode of the mechanical
resonator is made sufficiently strong, whereas the coupling to other degrees
of freedom in the environment is sufficiently weak. Satisfying this condition
experimentally when the interfering degree of freedom is a single electron, as
in the Ref. \cite{Armour_035311}, or a single photon, as in Refs.
\cite{Bose_4175,Mancini_3042,Marshall_130401,Bernad_0604157}, turns out to be
very difficult.

In the present paper we study an alternative configuration consisting of an RF
superconducting quantum interference device (SQUID) integrated with a
mechanical resonator in the shape of a doubly clamped beam. The dependence of
the total magnetic flux threading the loop on the beam's displacement leads to
a coupling between the RF SQUID and the mechanical resonator. We study the
effect of such a coupling on the visibility of Rabi oscillations between the
two lowest energy states of the RF SQUID, and discuss the required conditions
for experimental observation of decoherence and recoherence originating from
the coupling to the mechanical resonator.

The paper is organized as follows. The Hamiltonian for the closed system is
obtained in section II. An adiabatic approximation is employed in section III
to simplify the equations of motion of the system by considering the
mechanical motion as slow in comparison with the faster dynamics of the RF
SQUID. Further simplification is achieved in section IV by taking into account
only the two lowest energy levels of the RF SQUID. In section V we calculate
the effect of the mechanical resonator on the visibility of Rabi oscillations
between these two energy levels. Corrections due to finite temperature and
mechanical damping are considered in sections VI and VII respectively. The
validity of the adiabatic approximation is examined in section VIII. A
numerical example is given in section IX and discussion and conclusions are
given in section X.

Similar systems consisting of a SQUID integrated with a nanomechanical
resonator have been recently studied theoretically. Zhou and Mizel have shown
that nonlinear coupling between a DC SQUID and a mechanical resonator can be
employed for producing squeezed states of the mechanical resonator
\cite{Zhou_0605017}. More recently, Xue \textit{et al}. have shown that a flux
qubit integrated with a nanomechanical resonator can form a cavity quantum
electrodynamics system in the strong coupling region \cite{Xue_0607180}.

\section{Hamiltonian of the Closed System}

\begin{figure}
[ptb]
\begin{center}
\includegraphics[
height=3.0658in,
width=3.4411in
]%
{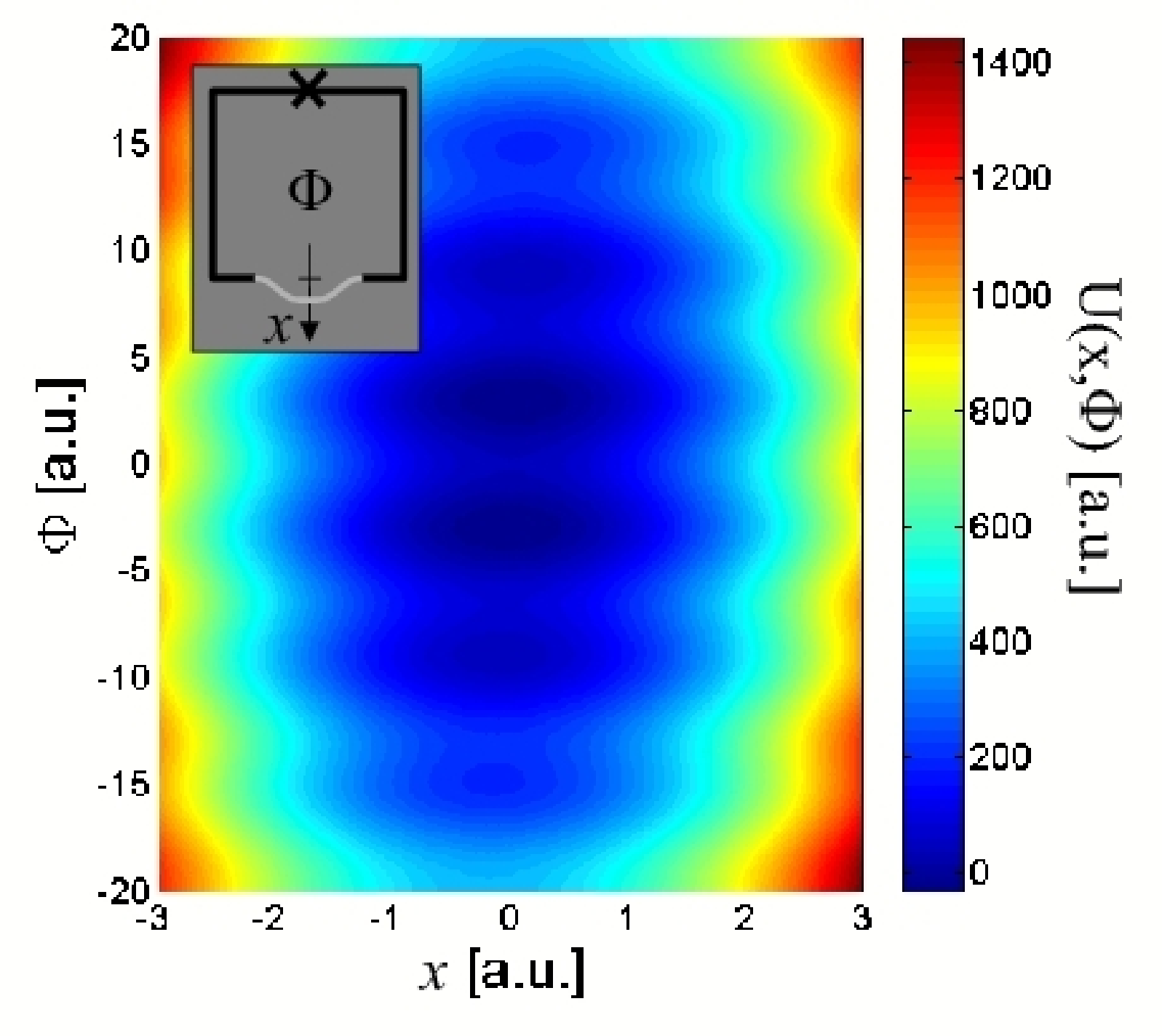}%
\caption{(Color online) The potential $U\left(  x,\Phi\right)  $ for the case
$\Phi_{e}=\Phi_{0}/2$ and $\beta_{L}=20$. The inset schematically shows the
device.}%
\label{U plot}%
\end{center}
\end{figure}

Consider the RF SQUID shown in the inset of Fig. \ref{U plot}, in which a
section of the loop is freely suspended and allowed to oscillate mechanically.
We assume the case where the fundamental mechanical mode vibrates in the plane
of the loop and denote the amplitude of this flexural mode as $x$. Let $m$ be
the effective mass of the fundamental mode, and $\omega_{0}$ its angular
frequency. A magnetic field is applied perpendicularly to the plane of the
loop. Let $\Phi_{e}$ be the externally applied flux for the case $x=0$, and
$B$ is the component of the magnetic field normal to the plane of the loop at
the location of the doubly clamped beam (it is assumed that $B$ is constant in
the region where the beam oscillates). The total magnetic flux $\Phi$
threading the loop is given by%

\begin{equation}
\Phi=\Phi_{e}+Blx+LI\ , \label{Phi}%
\end{equation}
where $L$ is the self inductance of the loop, and $l$ is an effective length
of the beam. The contribution of other mechanical modes of the beam to $\Phi$
is assumed to be negligibly small.%

%TCIMACRO{\FRAME{ftbpFU}{3.4411in}{3.0658in}{0pt}{\Qcb{(Color online) The
%potential $U\left(  x,\Phi\right)  $ for the case $\Phi_{e}=\Phi_{0}/2$ and
%$\beta_{L}=20$. The inset schematically shows the device.}}{\Qlb{U plot}%
%}{fig1.eps}{\special{ language "Scientific Word";  type "GRAPHIC";
%width 3.4411in;  height 3.0658in;  depth 0pt;  original-width 6.3737in;
%original-height 5.6749in;  cropleft "0";  croptop "1";  cropright "1";
%cropbottom "0";  filename 'Fig1.eps';file-properties "XNPEU";}}}%
%BeginExpansion
%EndExpansion

A Josephson junction (JJ) having a critical current $I_{c}$ and capacitance
$C$ is integrated into the loop. We first consider the dynamics of the closed
system consisting of the RF SQUID with the integrated doubly clamped beam. The
effect of damping due to coupling to other degrees of freedom in the
environment will be discussed later.

\subsection{Lagrangian}

The Lagrangian of the closed system is a function of the position $x$, flux
$\Phi$ and their time derivatives (denoted by overdot):%

\begin{equation}
\mathcal{L}=\frac{1}{2}m\dot{x}^{2}+\frac{C\dot{\Phi}^{2}}{2}-U\left(
x,\Phi\right)  \ ,
\end{equation}
where the potential energy $U$ is given by%

\begin{equation}
U=\frac{m\omega_{0}^{2}x^{2}}{2}+\frac{\left(  \Phi-\Phi_{e}-Blx\right)  ^{2}%
}{2L}-\frac{\Phi_{0}I_{c}\cos\left(  \frac{2\pi\Phi}{\Phi_{0}}\right)  }{2\pi
}\ ,
\end{equation}
and $\Phi_{0}=h/2e$ is the flux quantum (see Fig. \ref{U plot}). The resulting
Euler - Lagrange equations are%

\begin{equation}
m\ddot{x}+m\omega_{0}^{2}x-\frac{Bl}{L}\left(  \Phi-\Phi_{e}-Blx\right)  =0\ ,
\end{equation}

\begin{equation}
C\ddot{\Phi}+\frac{\Phi-\Phi_{e}-Blx}{L}+I_{c}\sin\left(  \frac{2\pi\Phi}%
{\Phi_{0}}\right)  =0\ .
\end{equation}
Note that the gauge invariant phase across the Josephson junction $\gamma_{J}$
is given by%

\begin{equation}
\gamma_{J}=2\pi n-\frac{2\pi\Phi}{\Phi_{0}}\ ,
\end{equation}
where $n$ is integer. By using this and Eq. (\ref{Phi}) the equations of
motion can be rewritten as%

\begin{equation}
m\ddot{x}+m\omega_{0}^{2}x-BlI=0\ , \label{x dot dot}%
\end{equation}

\begin{equation}
I_{c}\sin\gamma_{J}+C\frac{\Phi_{0}}{2\pi}\ddot{\gamma}_{J}=I\ .
\label{gamma dor dot}%
\end{equation}

The interpretation of these equations is straightforward. Eq. (\ref{x dot dot}%
) expresses Newton's 2nd law where the force is composed of the restoring
mechanical force and the Lorentz force acting on the movable beam. \ Whereas
Eq. (\ref{gamma dor dot}) states that the circulating current $I$ equals the
sum of the current $I_{c}\sin\gamma_{J}$ through the JJ and the current
$C\dot{V}$ through the capacitor, where the voltage $V$ is given by the second
Josephson equation $V=\left(  \Phi_{0}/2\pi\right)  \dot{\gamma}_{J}$.

\subsection{Hamiltonian}

The variables canonically conjugate to $x$ and $\Phi$ are $p=\partial
\mathcal{L}/\partial\dot{x}=m\dot{x}$ and $Q=\partial\mathcal{L}/\partial
\dot{\Phi}=C\dot{\Phi}$ respectively. The Hamiltonian is given by%

\begin{align}
\mathcal{H}  &  =\frac{p^{2}}{2m}+\frac{Q^{2}}{2C}+U\left(  x,\Phi\right)
\ .\nonumber\\
&  \label{Hamiltonian}%
\end{align}

Quantization is achieved by regarding the variables $x$, $p$, $\Phi$ and $Q$
as Hermitian operators satisfying the following commutation relations $\left[
x,p\right]  =\left[  \Phi,Q\right]  =i\hbar$ and $\left[  x,\Phi\right]
=\left[  x,Q\right]  =\left[  p,\Phi\right]  =\left[  p,Q\right]  =0$.

\section{Adiabatic Case}

The Hamiltonian (\ref{Hamiltonian}) can be written as $\mathcal{H}%
=\mathcal{H}_{0}+\mathcal{H}_{1}$, where%

\begin{equation}
\mathcal{H}_{0}=\frac{p^{2}}{2m}+\frac{1}{2}m\omega_{0}^{2}x^{2}\ ,
\label{Ham_0}%
\end{equation}

\begin{equation}
\mathcal{H}_{1}=\frac{Q^{2}}{2C}+u\left(  x,\Phi\right)  \ . \label{Ham_1}%
\end{equation}
Using the notation $U_{0}=\Phi_{0}^{2}/8\pi^{2}L$, $x_{\phi}=\Phi_{0}/Bl$, and
$\beta_{L}=2\pi LI_{c}/\Phi_{0}$, the term $u$ can be written as%

\begin{equation}
u=U_{0}\left[  4\pi^{2}\left(  \frac{\Phi-\Phi_{e}}{\Phi_{0}}-\frac{x}%
{x_{\phi}}\right)  ^{2}-2\beta_{L}\cos\left(  \frac{2\pi\Phi}{\Phi_{0}%
}\right)  \right]  \ . \label{u(x,Phi)}%
\end{equation}

As a basis for expanding the general state of the system we use the solutions
of the following Schr\"{o}dinger equation%

\begin{equation}
\mathcal{H}_{1}\left\vert n\left(  x\right)  \right\rangle =\varepsilon
_{n}\left(  x\right)  \left\vert n\left(  x\right)  \right\rangle \ ,
\label{e.v H_1}%
\end{equation}
where $x$ is treated here as a parameter (rather than a degree of freedom).
The local eigenvectors are assumed to be orthonormal%

\begin{equation}
\left\langle m\left(  x\right)  |n\left(  x\right)  \right\rangle =\delta
_{nm}\ .
\end{equation}

The wavefunctions associated with the local eigenstates%

\begin{equation}
\varphi_{n,x}\left(  \Phi^{\prime}\right)  =\left\langle \Phi^{\prime
}|n\left(  x\right)  \right\rangle \ ,
\end{equation}
are the solutions of the Schr\"{o}dinger equation%

\begin{equation}
\left[  -\frac{\hbar^{2}}{2C}\frac{\partial^{2}}{\partial\Phi^{2}}+u\left(
\Phi;x\right)  \right]  \varphi_{n,x}=\varepsilon_{n}\left(  x\right)
\varphi_{n,x}\ . \label{Schro Phi}%
\end{equation}
The total wave function is expanded as%

\begin{equation}
\psi\left(  x,\Phi,t\right)  =\sum_{n}\xi_{n}\left(  x,t\right)  \left\vert
n\left(  x\right)  \right\rangle \ . \label{psi_adi}%
\end{equation}

In the adiabatic approximation \cite{Moody_160} the time evolution of the
coefficients $\xi_{n}$ is governed by the following set of decoupled equations
of motion%

\begin{equation}
\left[  \frac{p^{2}}{2m}+V_{m}\left(  x\right)  \right]  \xi_{m}=i\hbar
\dot{\xi}_{m}\ , \label{Schro_adi}%
\end{equation}
where the adiabatic potentials $V_{m}\left(  x\right)  $ are given by%

\begin{equation}
V_{m}\left(  x\right)  =\frac{1}{2}m\omega_{0}^{2}x^{2}+\varepsilon_{m}\left(
x\right)  \ . \label{v_m(x)}%
\end{equation}
The validity of the adiabatic approximation will be discussed in section VIII below.

To numerically evaluate the eigenvalues $\varepsilon_{m}\left(  x\right)  $,
it is convenient to introduce the dimensionless variables $2\pi\Phi/\Phi
_{0}=\pi+\phi$, $2\pi\Phi_{e}/\Phi_{0}=\pi+\phi_{e}$, $2\pi x/x_{\phi}%
=\phi_{x}$. Using this notation the Schr\"{o}dinger equation (\ref{Schro Phi})
can be rewritten as%

\begin{equation}
\left(  -\beta_{C}\frac{\partial^{2}}{\partial\phi^{2}}+\frac{u}{U_{0}%
}\right)  \varphi_{n,x}=\lambda_{n,x}\varphi_{n,x}\ , \label{Scrodinger phi}%
\end{equation}
where $\beta_{C}=2e^{2}/CU_{0}$,$\lambda_{n,x}=\varepsilon_{n}\left(
x\right)  /U_{0}$, and
\begin{equation}
\frac{u}{U_{0}}=\left(  \phi-\phi_{e}-\phi_{x}\right)  ^{2}+2\beta_{L}\cos
\phi\ . \label{u}%
\end{equation}

\section{Two-Level Approximation}

Consider the case where $\left\vert \phi_{e}\right\vert \ll1$ (namely,
$\Phi_{e}\simeq\Phi_{0}/2$), $\left\vert \phi_{x}\right\vert \ll1,$ and
$\beta_{L}>1$. In this case the local potential $u\left(  \phi\right)  $ given
by Eq. (\ref{u}) contains two wells separated by a barrier near $\phi=0$. At
low temperatures only the two lowest energy levels contribute. In this limit
the local Hamiltonian $\mathcal{H}_{1}$ can be expressed in the basis of the
states $\left\vert \curvearrowleft\right\rangle $ and $\left\vert
\curvearrowright\right\rangle $, representing localized states in the left and
right well respectively having opposite circulating currents. In this basis,
$\mathcal{H}_{1}$ is represented by the $2\times2$ matrix%

\begin{equation}
\mathcal{H}_{1}=\left(
\begin{array}
[c]{cc}%
\eta\left(  \phi_{e}+\phi_{x}\right)  & \Delta\\
\Delta & -\eta\left(  \phi_{e}+\phi_{x}\right)
\end{array}
\right)  \ . \label{2-level H_1}%
\end{equation}

The real parameters $\eta$ and $\Delta$ can be determined by solving
numerically the Schr\"{o}dinger equation (\ref{Scrodinger phi}).

Using the notation%

\begin{equation}
\tan\theta=\frac{\Delta}{\eta\left(  \phi_{e}+\phi_{x}\right)  }\ ,
\end{equation}
$\mathcal{H}_{1}$ can be rewritten as%

\begin{equation}
\mathcal{H}_{1}=\sqrt{\eta^{2}\left(  \phi_{e}+\phi_{x}\right)  ^{2}%
+\Delta^{2}}\left(
\begin{array}
[c]{cc}%
\cos\theta & \sin\theta\\
\sin\theta & -\cos\theta
\end{array}
\right)  \ .
\end{equation}

The eigenvectors and eigenenergies are denoted as%

\begin{equation}
\mathcal{H}_{1}\left\vert \pm\right\rangle =\varepsilon_{\pm}\left\vert
\pm\right\rangle \ ,
\end{equation}
where%

\begin{equation}
\left\vert +\right\rangle =\left(
\begin{array}
[c]{c}%
\cos\frac{\theta}{2}\\
\sin\frac{\theta}{2}%
\end{array}
\right)  ;\ \left\vert -\right\rangle =\left(
\begin{array}
[c]{c}%
-\sin\frac{\theta}{2}\\
\cos\frac{\theta}{2}%
\end{array}
\right)  \ ,
\end{equation}

\begin{equation}
\varepsilon_{\pm}=\pm\sqrt{\eta^{2}\left(  \phi_{e}+\phi_{x}\right)
^{2}+\Delta^{2}}\ .
\end{equation}

\section{Rabi Oscillations}

Consider the following experimental protocol for detecting Rabi oscillations
between the two lowest energy states of the RF SQUID. The first stage consists
of state preparation performed by applying a large constant external flux
$\phi_{e}$. At time $t=0$ the external flux is switched off and the system
starts oscillating. At a later time $t>0$ the final state of the RF SQUID is measured.

\subsection{State Preparation}

The system is first prepared in an initial state by applying an external bias
flux $\phi_{e}$ such that $\phi_{e}\gg\Delta/\eta$. In this limit one finds
approximately $\left\vert +\right\rangle =\left\vert \curvearrowleft
\right\rangle $, $\left\vert -\right\rangle =\left\vert \curvearrowright
\right\rangle $, and $\varepsilon_{\pm}=\pm\eta\left(  \phi_{e}+\phi
_{x}\right)  $. Thus, the adiabatic potentials Eq. (\ref{v_m(x)}) are given by%

\begin{align}
V_{\pm}\left(  x\right)   &  =\frac{1}{2}m\omega_{0}^{2}\left(  x\pm
x_{0}\right)  ^{2}-\eta\left(  \frac{\pi x_{0}}{x_{\phi}}\mp\phi_{e}\right)
\ ,\nonumber\\
&  \label{V up down small Delta}%
\end{align}
where%

\begin{equation}
x_{0}=\frac{2\pi\eta}{m\omega_{0}^{2}x_{\phi}}\ . \label{x_0}%
\end{equation}

Assume also the case where the temperature $T$ is relatively low $k_{B}%
T\ll\Delta$. In this limit the RF SQUID is expected to occupy its ground state
$\left\vert \curvearrowright\right\rangle $ in thermal equilibrium. The
mechanical resonator is expected to be in a thermal state of the potential
well $V_{-}\left(  x\right)  $ centered at $x_{0}$ [see Eq.
(\ref{V up down small Delta})].

\subsection{Switching off the External Flux}

At time $t=0$, the external flux $\phi_{e}$ is suddenly switched to a new
value $\phi_{e}=0$. Using the notation%

\begin{equation}
\zeta=\frac{\eta}{\Delta}\frac{2\pi x_{0}}{x_{\phi}}=\frac{m\omega_{0}%
^{2}x_{0}^{2}}{\Delta}\ , \label{zeta}%
\end{equation}
one finds to lowest order in $\phi_{x}$%

\begin{align}
\left\vert \pm\right\rangle  &  =\frac{\sqrt{2}}{2}\left(
\begin{array}
[c]{c}%
1\pm\frac{\zeta}{2}\frac{x}{x_{0}}\\
\pm1-\frac{\zeta}{2}\frac{x}{x_{0}}%
\end{array}
\right)  \ ,\label{up down small phi}\\
\varepsilon_{\pm}  &  =\pm\Delta\left(  1+\frac{\zeta^{2}}{2}\frac{x^{2}%
}{x_{0}^{2}}\right)  \ , \label{epsilon up down}%
\end{align}
and the adiabatic potentials (\ref{v_m(x)}) for this case are given by%

\begin{equation}
V_{\pm}\left(  x\right)  =\frac{1}{2}m\omega_{0}^{2}\left(  1\pm\zeta\right)
x^{2}\pm\Delta\ . \label{V up down}%
\end{equation}

Thus, both mechanical states associated with the RF SQUID states $\left\vert
+\right\rangle $ and $\left\vert -\right\rangle $ will at $t=0$ start
oscillating with different frequencies $\omega_{0}\sqrt{1+\zeta}$ and
$\omega_{0}\sqrt{1-\zeta}$ respectively around the point $x=0$. Consider the
case where $\zeta\ll1$. Using Eq. (\ref{up down small phi}) one finds that the approximation%

\begin{equation}
\left\vert \pm\left(  x\right)  \right\rangle =\left\vert \pm\left(
x=0\right)  \right\rangle =\frac{\sqrt{2}}{2}\left(
\begin{array}
[c]{c}%
1\\
\pm1
\end{array}
\right)  \ , \label{up down small zeta}%
\end{equation}
can be employed in the region $\left\vert x\right\vert \lesssim x_{0}$ where
the mechanical resonator oscillates.

\subsection{Measuring the RF SQUID Final State}

Consider the case where the mechanical system was at time $t=0$ in a given
state, denoted as $\left\vert \xi_{0}\right\rangle _{e}$, with a wave function
$\xi_{0}\left(  x\right)  $. We first calculate the time evolution for a given
state, and later perform a thermal averaging over initial states. The state of
the system at $t=0$ can be expressed as%

\begin{equation}
\psi\left(  t=0\right)  =\xi_{+}\left(  t=0\right)  \left\vert +\left(
x=0\right)  \right\rangle +\xi_{-}\left(  t=0\right)  \left\vert -\left(
x=0\right)  \right\rangle \ ,
\end{equation}
where%

\begin{equation}
\xi_{\pm}\left(  t=0\right)  =\pm\frac{\sqrt{2}}{2}\xi_{0}\left(  x\right)
\ .
\end{equation}

In the last step the state of the RF SQUID is measured. What is the
probability to find the RF SQUID in a given state $\left\vert \chi
\right\rangle $ at time $t$ ? To calculate this probability $P_{\left\vert
\chi\right\rangle }\left(  t\right)  $ one has to trace out the mechanical
degree of freedom. By using Eq. (\ref{psi_adi}) and employing the two-level
approximation one finds in general%

\begin{align}
P_{\left\vert \chi\right\rangle }\left(  t\right)   &  =\int dx\left\vert
\xi_{+}\left(  x,t\right)  \left\langle \chi|+\left(  x\right)  \right\rangle
+\xi_{-}\left(  x,t\right)  \left\langle \chi|-\left(  x\right)  \right\rangle
\right\vert ^{2}\ .\nonumber\\
&
\end{align}

As an example, consider the case where $\left\vert \chi\right\rangle
=\left\vert \curvearrowright\right\rangle $. Using Eq.
(\ref{up down small zeta}) one finds%

\begin{equation}
P_{\left\vert \curvearrowright\right\rangle }\left(  t\right)  =\frac{1}%
{2}+\operatorname{Re}\int dx\xi_{+}\left(  x,t\right)  \xi_{-}^{\ast}\left(
x,t\right)  \ .
\end{equation}
Alternatively, using Eqs. (\ref{Schro_adi}) and (\ref{V up down}) this can be
expressed as%

\begin{align}
P_{\left\vert \curvearrowright\right\rangle }\left(  t\right)   &  =\frac
{1}{2}+\frac{1}{2}\operatorname{Re}\ \left[  \nu_{0}\left(  t\right)
\exp\left(  -\frac{2i\Delta t}{\hbar}\right)  \right]  \ ,\nonumber\\
&  \label{P_|ccw>}%
\end{align}
where%

\begin{equation}
\nu_{0}\left(  t\right)  =\ _{e}\left\langle \xi_{0}\right\vert \exp\left(
\frac{iH\left(  -\zeta\right)  t}{\hbar}\right)  \exp\left(  -\frac{iH\left(
\zeta\right)  t}{\hbar}\right)  \left\vert \xi_{0}\right\rangle _{e},
\end{equation}
and%

\begin{equation}
H\left(  \zeta\right)  =\frac{p^{2}}{2m}+\frac{1}{2}m\omega_{0}^{2}\left(
1+\zeta\right)  x^{2}\ .
\end{equation}
Eq. (\ref{P_|ccw>}) indicates that the visibility of Rabi oscillations
(occurring at angular frequency $2\Delta/\hbar$) is diminished by the factor
$\left\vert \nu_{0}\left(  t\right)  \right\vert $ (note that in general
$\left\vert \nu_{0}\left(  t\right)  \right\vert \leq1$).

The Hamiltonian $H_{\zeta}$ can be written as%

\begin{equation}
H\left(  \zeta\right)  =\sqrt{1+\zeta}H\left(  0\right)  +V_{\zeta}\ ,
\end{equation}
where%

\begin{equation}
V_{\zeta}=\frac{p^{2}}{2m}\left(  1-\sqrt{1+\zeta}\right)  +\frac{m\omega
_{0}^{2}x^{2}}{2}\left(  1-\sqrt{1+\zeta}+\zeta\right)  \ .
\end{equation}

The Hamiltonian $\sqrt{1+\zeta}H\left(  0\right)  $ is associated with a
harmonic oscillator having mass $m/\sqrt{1+\zeta}$ and a resonance frequency
$\omega_{0}\sqrt{1+\zeta}$. Assuming that $\zeta\ll1$ one can employ the approximation%

\begin{equation}
H\left(  \zeta\right)  \simeq\sqrt{1+\zeta}H\left(  0\right)  \ .
\label{H(zeta)}%
\end{equation}
This approximation greatly simplifies the analysis since annihilation and
creation operators associated with both Hamiltonians $H\left(  \zeta\right)  $
and $H\left(  -\zeta\right)  $ are common. Note that the time evolution
generated by both Hamiltonians, $H\left(  \zeta\right)  \ $and $\sqrt{1+\zeta
}H\left(  0\right)  $, is periodic in time with the same period $2\pi
/\omega_{0}\sqrt{1+\zeta}$. Thus, the error introduced by this approximation
is small even for times much longer than the period time, provided that the
condition $\zeta\ll1$ is satisfied. Using this approximation and keeping terms
up to first order in $\zeta$ yield%

\begin{equation}
\nu_{0}\left(  t\right)  =\ _{e}\left\langle \xi_{0}\right\vert \exp\left(
-\frac{i\zeta H\left(  0\right)  t}{\hbar}\right)  \left\vert \xi
_{0}\right\rangle _{e}\ .
\end{equation}

\section{Thermal Averaging}

At finite temperature $T$ the term $\nu_{0}\left(  t\right)  $ has to be
calculated by averaging over a thermal distribution of initial states
$\left\vert \xi_{0}\right\rangle _{e}$. At times $t<0$ the mechanical
resonator is expected to be in a thermal state of the potential well
$V_{-}\left(  x\right)  $ centered at $x_{0}$ [Eq.
(\ref{V up down small Delta})]. It is convenient to express this thermal
distribution using a displacement operator $D\left(  \alpha_{0}\right)  $, where%

\begin{equation}
D\left(  \alpha\right)  =\exp\left[  \sqrt{\frac{m\omega_{0}}{2\hbar}}\left(
\alpha-\alpha^{\ast}\right)  x-i\sqrt{\frac{1}{2\hbar m\omega_{0}}}\left(
\alpha+\alpha^{\ast}\right)  p\right]  \ ,
\end{equation}
and%

\begin{equation}
\alpha_{0}=x_{0}\sqrt{\frac{m\omega_{0}}{2\hbar}}\ .
\end{equation}

For a general c-number $\alpha$, the operator $D\left(  \alpha\right)  $
transforms the vacuum state $\left\vert 0\right\rangle $ into a coherent state
$\left\vert \alpha\right\rangle $, i.e., $D\left(  \alpha\right)  \left\vert
0\right\rangle =\left\vert \alpha\right\rangle $. Using this notation one finds%

\begin{equation}
\nu_{0}\left(  t\right)  =\ \left\langle D^{\dag}\left(  \alpha_{0}\right)
\exp\left(  -\frac{i\zeta H\left(  0\right)  t}{\hbar}\right)  D\left(
\alpha_{0}\right)  \right\rangle \ , \label{nu_0=<>}%
\end{equation}
where the brackets $\left\langle {}\right\rangle $ represent thermal
averaging. It is convenient to employ the coherent states diagonal
representation (\textit{P }representation) \cite{Glauber_QO} of the density
operator at thermal equilibrium%

\begin{equation}
\rho=\int\int\mathrm{d}^{2}\alpha P\left(  \alpha\right)  \left\vert
\alpha\right\rangle \left\langle \alpha\right\vert \ ,
\end{equation}
where $\mathrm{d}^{2}\alpha$ denotes infinitesimal area in the $\alpha$
complex plane, namely $\mathrm{d}^{2}\alpha=\mathrm{d}\left\{
\operatorname{Re}\alpha\right\}  \mathrm{d}\left\{  \operatorname*{Im}%
\alpha\right\}  $, the probability density $P\left(  \alpha\right)  $ is given by%

\begin{equation}
P\left(  \alpha\right)  =\frac{1}{\pi\left\langle n\right\rangle }\exp\left(
-\frac{\left\vert \alpha\right\vert ^{2}}{\left\langle n\right\rangle
}\right)  \ ,
\end{equation}
and%

\begin{equation}
\left\langle n\right\rangle =\frac{1}{e^{\hbar\omega_{0}/k_{B}T}-1}\ ,
\end{equation}
is the thermal occupation number.

Thus%

\begin{align}
\nu_{0}\left(  t\right)   &  =\operatorname{Tr}\left[  \rho D^{\dag}\left(
\alpha_{0}\right)  \exp\left(  -\frac{i\zeta H\left(  0\right)  t}{\hbar
}\right)  D\left(  \alpha_{0}\right)  \right] \nonumber\\
&  =\int\int\mathrm{d}^{2}\alpha P\left(  \alpha\right) \nonumber\\
&  \times\left\langle \alpha\right\vert D^{\dag}\left(  \alpha_{0}\right)
\exp\left(  -\frac{i\zeta H\left(  0\right)  t}{\hbar}\right)  D\left(
\alpha_{0}\right)  \left\vert \alpha\right\rangle \ .\nonumber\\
&  \label{nu_0=Tr}%
\end{align}

Using the identity%

\begin{equation}
D\left(  \alpha_{0}\right)  \left\vert \alpha\right\rangle =\exp\left(
\frac{\alpha_{0}\alpha^{\ast}-\alpha_{0}^{\ast}\alpha}{2}\right)  \left\vert
\alpha_{0}+\alpha\right\rangle \ , \label{D(alpha_0)|alpha>}%
\end{equation}
and noting that $\alpha_{0}$ is real yield%

\begin{align}
\nu_{0}\left(  t\right)   &  =\frac{e^{-i\zeta\omega_{0}t/2}}{\pi\left\langle
n\right\rangle }\int\int\mathrm{d}^{2}\alpha\ \exp\left(  -\frac{\left\vert
\alpha\right\vert ^{2}+\xi\left\vert \alpha_{0}+\alpha\right\vert ^{2}%
}{\left\langle n\right\rangle }\right) \nonumber\\
&  =\frac{e^{-i\zeta\omega_{0}t/2}}{\pi}\exp\left(  -\frac{\xi a_{0}^{2}%
}{1+\xi}\right) \nonumber\\
&  \times\int_{-\infty}^{\infty}\mathrm{d}x\exp\left[  -\left(  1+\xi\right)
\left(  x+\frac{\xi a_{0}}{1+\xi}\right)  ^{2}\right] \nonumber\\
&  \times\int_{-\infty}^{\infty}\mathrm{d}y\ \exp\left[  -\left(
1+\xi\right)  y^{2}\right]  \ ,\nonumber\\
&  \label{nu_0=}%
\end{align}
where $\xi=\left(  1-e^{-i\zeta\omega_{0}t}\right)  \left\langle
n\right\rangle $ and $a_{0}=\alpha_{0}/\sqrt{\left\langle n\right\rangle }$.

In the limit of zero temperature where $\left\langle n\right\rangle
\rightarrow0$ one finds%

\begin{equation}
\nu_{0}\left(  t\right)  =e^{-i\zeta\omega_{0}t/2}\exp\left[  -\alpha_{0}%
^{2}\left(  1-e^{-i\zeta\omega_{0}t}\right)  \right]  \ ,
\end{equation}
and the visibility factor in this limit is given by%

\begin{equation}
\left\vert \nu_{0}\left(  t\right)  \right\vert ^{2}=\exp\left[  -4\alpha
_{0}^{2}\sin^{2}\left(  \frac{\zeta\omega_{0}t}{2}\right)  \right]  \ .
\end{equation}

Another case of interest is the limit of short times. The term $\left\vert
\nu_{0}\left(  t\right)  \right\vert ^{2}$ is calculated to lowest order in
$t$ using Eq. (\ref{nu_0=<>}) and perturbation theory%

\begin{equation}
\left\vert \nu_{0}\left(  t\right)  \right\vert ^{2}=1-\left(  \frac{\zeta
t}{\hbar}\right)  ^{2}V_{H}\ ,
\end{equation}
where%

\begin{equation}
V_{H}=\left\langle D^{\dag}\left(  \alpha_{0}\right)  H^{2}\left(  0\right)
D\left(  \alpha_{0}\right)  \right\rangle -\left\langle D^{\dag}\left(
\alpha_{0}\right)  H\left(  0\right)  D\left(  \alpha_{0}\right)
\right\rangle ^{2}\ .
\end{equation}
Using Eqs. (\ref{nu_0=Tr}) and (\ref{D(alpha_0)|alpha>}) one finds%

\begin{equation}
V_{H}=\hbar^{2}\omega_{0}^{2}\left(  \alpha_{0}^{2}+\left\langle
n\right\rangle \right)  \ .
\end{equation}
The result can be expressed in terms of a decoherence rate $\gamma_{m}$%
\begin{equation}
\left\vert \nu_{0}\left(  t\right)  \right\vert ^{2}=1-\left(  \gamma
_{m}t\right)  ^{2}\ ,
\end{equation}
where%

\begin{equation}
\gamma_{m}=\zeta\alpha_{0}\omega_{0}\left(  1+\frac{\left\langle
n\right\rangle }{\alpha_{0}^{2}}\right)  ^{1/2}\ .
\end{equation}

\section{Effect of Mechanical Damping}

Consider in general a mechanical resonator in a superposition of two coherent
states $\left\vert \alpha_{1}\right\rangle $ and $\left\vert \alpha
_{2}\right\rangle $. Coupling between the resonator and a thermal bath at
temperature $T$ induces decoherence with a rate $\gamma_{d}$ given by
\cite{Caldeira_587, Joos_223, Unruh_1071, Zurek_36}%

\begin{equation}
\gamma_{d}=\frac{2\omega_{0}}{Q}\left\vert \alpha_{1}-\alpha_{2}\right\vert
^{2}\coth\frac{\hbar\omega_{0}}{2k_{B}T}\ ,
\end{equation}
where $\omega_{0}$ and $Q$ are the resonance frequency and quality factor respectively.

Damping is thus expected to further diminish the visibility of Rabi
oscillations. The factor $\nu\left(  t\right)  $ is written as%

\begin{equation}
\nu\left(  t\right)  =\nu_{0}\left(  t\right)  \nu_{d}\left(  t\right)  \ ,
\end{equation}
where $\nu_{d}\left(  t\right)  $ represents the contribution of damping.

To provide a rough estimate of the factor $\nu_{d}\left(  t\right)  $ in the
present case the c-numbers $\alpha_{1}$ and $\alpha_{2}$ are substituted by
the thermal average values of the distributions associated with the
$\left\vert +\right\rangle $ and $\left\vert -\right\rangle $ states
respectively \cite{Armour_035311}, and thus we take%

\begin{subequations}
\begin{align}
\alpha_{1}\left(  t\right)   &  =\alpha_{0}\exp\left[  -i\left(  1+\frac
{\zeta}{2}\right)  \omega_{0}t\right]  \ ,\\
\alpha_{2}\left(  t\right)   &  =\alpha_{0}\exp\left[  -i\left(  1-\frac
{\zeta}{2}\right)  \omega_{0}t\right]  \ .
\end{align}

We further require that%

\end{subequations}
\begin{equation}
\frac{\mathrm{d}\nu_{d}}{\mathrm{d}t}=-\gamma_{d}\nu_{d}\ ,
\end{equation}
and obtain%

\begin{align}
\nu_{d}\left(  t\right)   &  =\exp\left[  -\frac{4\alpha_{0}^{2}\omega_{0}%
t}{Q}\coth\frac{\hbar\omega_{0}}{2k_{B}T}\left(  1-\frac{\sin\left(
\zeta\omega_{0}t\right)  }{\zeta\omega_{0}t}\right)  \right]  \ .\nonumber\\
&
\end{align}

Recall that the recoherence peaks, where $\left\vert \nu_{0}\left(
t_{n}\right)  \right\vert =1$, occur at times $t_{n}=2\pi n/\zeta\omega_{0}$,
where $n$ is integer [see Eq. (\ref{nu_0=})]. Recoherence can be detected only
if $\gamma_{d}$ is sufficiently small. For the first recoherence peak at time
$t_{1}$, we have%

\begin{equation}
\nu_{d}\left(  t_{1}\right)  =\exp\left[  -\frac{8\pi\alpha_{0}^{2}}{\zeta
Q}\coth\frac{\hbar\omega_{0}}{2k_{B}T}\right]  \ ,
\end{equation}
whereas for the other recoherence peaks the following holds%

\begin{equation}
\nu_{d}\left(  t_{n}\right)  =\left[  \nu_{d}\left(  t_{1}\right)  \right]
^{n}\ .
\end{equation}

In the case $\hbar\omega_{0}\ll k_{B}T$ one has%

\begin{equation}
\nu_{d}\left(  t_{1}\right)  =\exp\left[  -\frac{4\pi}{\zeta Q}\left(
\frac{x_{0}}{\lambda_{T}}\right)  ^{2}\right]  \ ,
\end{equation}
where $\lambda_{T}$ is the thermal length%

\begin{equation}
\lambda_{T}=\frac{\hbar}{\sqrt{2mk_{B}T}}\ .
\end{equation}

\section{Adiabatic Condition}

We now return to the adiabatic approximation and examine its validity. In the
adiabatic limit the off-diagonal terms in the set of coupled equations for the
amplitudes $\xi_{n}$ are considered negligibly small, and consequently no
Zener transitions between adiabatic states occur. This approximation yields
the set of decoupled equations (\ref{Schro_adi}). To calculate the Zener
transition probability to lowest order we consider the off diagonal elements
as a perturbation.

Consider mechanical oscillations with an amplitude $x_{0}$ and assume the case
where $\phi_{e}=0$. A Zener transition is most likely to occur near the times
when the mechanical resonator crosses the point $x=0$, namely, when the
mechanical velocity peaks and the energy gap $\varepsilon_{+}-\varepsilon_{-}$
obtains its smallest value. The probability $p_{Z}$ that a Zener transition
will occur per such a crossing can be calculated using Eq. (C25) of Ref.
\cite{Buks_628}%
\begin{equation}
p_{Z}=\exp\left(  -\frac{x_{\phi}}{2x_{0}\beta_{h}}\frac{\Delta^{2}}{\eta
U_{0}}\right)  \ ,
\end{equation}
where $\beta_{h}=\hbar\omega_{0}/U_{0}$. The adiabatic approximation is valid
when $p_{Z}\ll1$.

\section{Estimation of Parameters}

Satisfying all the above mentioned conditions required for experimental
observation of decoherence and recoherence is quite challenging. However, a
careful design together with state of the art fabrication and cryogenics
techniques may allow experimental implementation. We examine below an example
of a device having the following parameters%

\begin{subequations}
\begin{align}
L  &  =6.5\times10^{-11}%
%TCIMACRO{\unit{H}}%
%BeginExpansion
\operatorname{H}%
%EndExpansion
\ ,\\
C  &  =7.4\times10^{-17}%
%TCIMACRO{\unit{F}}%
%BeginExpansion
\operatorname{F}%
%EndExpansion
\ ,\\
I_{c}  &  =10%
%TCIMACRO{\unit{\U{3bc}A}}%
%BeginExpansion
\operatorname{\mu A}%
%EndExpansion
\ ,\\
m  &  =10^{-16}%
%TCIMACRO{\unit{kg}}%
%BeginExpansion
\operatorname{kg}%
%EndExpansion
\ ,\\
\omega_{0}/2\pi &  =640%
%TCIMACRO{\unit{MHz}}%
%BeginExpansion
\operatorname{MHz}%
%EndExpansion
\ ,\\
Bl  &  =%
%TCIMACRO{\unit{T}}%
%BeginExpansion
\operatorname{T}%
%EndExpansion
\times%
%TCIMACRO{\unit{\U{3bc}m}}%
%BeginExpansion
\operatorname{\mu m}%
%EndExpansion
\ ,\\
Q  &  =10^{4}\ ,\\
T  &  =0.05%
%TCIMACRO{\unit{K}}%
%BeginExpansion
\operatorname{K}%
%EndExpansion
\ .
\end{align}
These parameters for both the RF SQUID \cite{Friedman_43} and for the
nanomechanical resonator \cite{Roukes_0008187} are within reach with present
day technology.

The chosen value of $L$ corresponds to a circular loop with a radius of about
$10%
%TCIMACRO{\unit{\U{3bc}m}}%
%BeginExpansion
\operatorname{\mu m}%
%EndExpansion
$ and a wire having a cross section of about $\left(  0.1%
%TCIMACRO{\unit{\U{3bc}m}}%
%BeginExpansion
\operatorname{\mu m}%
%EndExpansion
\right)  ^{2}$, whereas the values of $C$ and $I_{c}$ correspond to a junction
having a plasma frequency of about $8%
%TCIMACRO{\unit{THz}}%
%BeginExpansion
\operatorname{THz}%
%EndExpansion
$. The parameter $Bl$ plays a crucial role in determining the coupling
strength between the mechanical resonator and the RF SQUID. Enhancing the
coupling can be achieved by increasing the applied magnetic field at the
location of the mechanical resonator $B$. However, $B$ should not exceed the
superconducting critical field. Moreover, the externally applied magnetic
field at the location of the JJ must be kept at a much lower value in order to
minimize an undesirable reduction in $I_{c}$. This can be achieved by
employing an appropriate design in which the applied field is strongly nonuniform.

Using these values one finds%

\end{subequations}
\begin{subequations}
\begin{align}
\beta_{L}  &  =1.9\ ,\\
\beta_{C}  &  =0.78\ ,\\
\beta_{h}  &  =4.8\times10^{-4}\ ,\\
\frac{U_{0}}{k_{B}}  &  =64%
%TCIMACRO{\unit{K}}%
%BeginExpansion
\operatorname{K}%
%EndExpansion
\ ,\\
x_{\phi}  &  =2.1%
%TCIMACRO{\unit{nm}}%
%BeginExpansion
\operatorname{nm}%
%EndExpansion
\ ,\\
x_{\phi}\sqrt{\frac{m\omega_{0}}{2\hbar}}  &  =9.1\times10^{4}\ ,\\
\lambda_{T}  &  =9.0\times10^{-6}%
%TCIMACRO{\unit{nm}}%
%BeginExpansion
\operatorname{nm}%
%EndExpansion
\ .
\end{align}

The values of $\beta_{L}$ and $\beta_{C}$ are employed for calculating
numerically the eigenstates of Eq. (\ref{Scrodinger phi}). \ Fig.
\ref{eigenstates} (a)-(c) shows the first 3 lowest energy states for the case
$\phi_{e}+\phi_{x}=0$, whereas panel (d) shows the dependence of the energy of
the two lowest energy states on $\phi_{e}+\phi_{x}$.%
%TCIMACRO{\FRAME{ftbpFU}{3.2396in}{4.2704in}{0pt}{\Qcb{(Color online)
%Eigenstates of $\mathcal{H}_{1}$. (a)-(c) The first 3 lowest energy states for
%the case $\phi_{e}+\phi_{x}=0$. (d) The energy of the two lowest states vs.
%$\phi_{e}+\phi_{x}$.}}{\Qlb{eigenstates}}{fig2.eps}%
%width 3.2396in;  height 4.2704in;  depth 0pt;  original-width 6.0079in;
%original-height 8.0851in;  cropleft "0";  croptop "1";  cropright "1";
%cropbottom "0";  filename 'Fig2.eps';file-properties "XNPEU";}}}%
%BeginExpansion
\begin{figure}
[ptb]
\begin{center}
\includegraphics[
height=4.2704in,
width=3.2396in
]%
{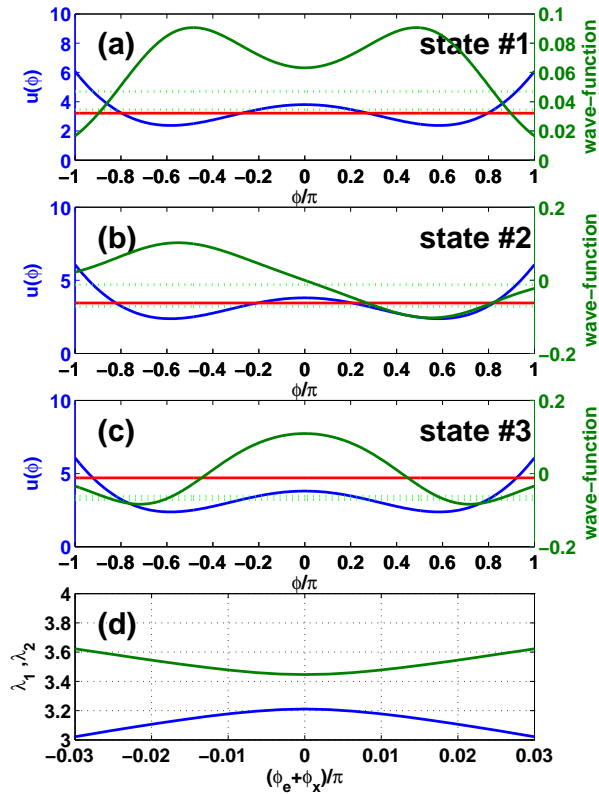}%
\caption{(Color online) Eigenstates of $\mathcal{H}_{1}$. (a)-(c) The first 3
lowest energy states for the case $\phi_{e}+\phi_{x}=0$. (d) The energy of the
two lowest states vs. $\phi_{e}+\phi_{x}$.}%
\label{eigenstates}%
\end{center}
\end{figure}
%EndExpansion

From these results one finds for the values of the $\eta$ and $\Delta$
parameters in the two-level approximation to Hamiltonian $\mathcal{H}_{1}$
[Eq. (\ref{2-level H_1})],%

\end{subequations}
\begin{subequations}
\begin{align}
\eta &  =2.5U_{0}\ ,\\
\Delta &  =0.12U_{0}\ .
\end{align}

Using these values yields%

\end{subequations}
\begin{subequations}
\begin{align}
x_{0}  &  =4.1\times10^{-6}%
%TCIMACRO{\unit{nm}}%
%BeginExpansion
\operatorname{nm}%
%EndExpansion
\ ,\\
\zeta &  =2.5\times10^{-4}\ ,\\
\frac{x_{0}}{x_{\phi}}  &  =2.0\times10^{-6}\ ,\\
\alpha_{0}  &  =0.18\ ,\\
\left\langle n\right\rangle  &  =1.2\ ,\\
\left(  \zeta\alpha_{0}\omega_{0}\right)  ^{-1}  &  =5.6%
%TCIMACRO{\unit{\U{3bc}s}}%
%BeginExpansion
\operatorname{\mu s}%
%EndExpansion
\ ,\label{(zeta alpha omega_0)^(-1)}\\
\frac{4\pi}{\zeta Q}\left(  \frac{x_{0}}{\lambda_{T}}\right)  ^{2}  &
=1.0\ ,\label{-4 pi / zeta Q}\\
\frac{2\pi}{\zeta\omega_{0}}  &  =6.3%
%TCIMACRO{\unit{\U{3bc}s}}%
%BeginExpansion
\operatorname{\mu s}%
%EndExpansion
\ ,\\
\frac{x_{\phi}}{2x_{0}\beta_{h}}\frac{\Delta^{2}}{\eta U_{0}}  &
=3.0\times10^{6}\ . \label{4 pi U_0^2}%
\end{align}

Eqs. (\ref{(zeta alpha omega_0)^(-1)}) and (\ref{-4 pi / zeta Q}) indicate
that observation of both decoherence and recoherence, for the case of the
present example, is feasible, provided that the decoherence time of the RF
SQUID due to other mechanisms is sufficiently long, i.e., on the order of
microseconds \cite{Yoshihara_0606481}. Moreover, Eq. (\ref{4 pi U_0^2})
ensures the validity of the adiabatic approximation.

\section{Discussion and Conclusions}

A possible, alternative protocol to the presently considered one for observing
decoherence/recoherence phenomena is the so-called Ramsey interference
experiment that proceeds as follows~\cite{Armour_148301} : (i) At time $t<0$,
the state is prepared in the ground state $\left\vert \curvearrowright
\right\rangle $, identically to the above considered protocol by applying an
external bias flux $\phi_{e}$ such that $\phi_{e}\gg\Delta/\eta$ ; (ii) At
time $t=0$, the external flux $\phi_{e}$ is suddenly switched to the new value
$\phi_{e}=0$, again just as in the above protocol, but then after one-quarter
of a Rabi oscillation period, $\phi_{e}$ is suddenly switched back up to the
same non-zero value as was applied during first, preparation stage; (iii) The
flux qubit and mechanical oscillator are then left to interact for a certain
duration with $\phi_{e}$ kept constant; (iv) Stage (ii) is repeated again; (v)
The state of the qubit is read out.

The effect of stage (ii) is to prepare the flux qubit in a state which is an
equal magnitude superposition of the circulating current states $\left\vert
\curvearrowright\right\rangle $ and $\left\vert \curvearrowleft\right\rangle
$. Each of these states is associated with the different spatially-shifted
potentials $V_{\pm}(x)$ [Eq. (\ref{V up down small Delta})], so that during
the interaction stage (iii) an entangled state develops between the oscillator
and flux qubit, giving rise to decoherence of the reduced qubit state. After
one full mechanical period, the entanglement is undone, resulting in
recoherence. The second, quarter Rabi period pulse enables one to probe the
decoherence/recoherence, simply by measuring the probability to be in one of
the measurement basis states, e.g., the ground state $\left\vert
\curvearrowright\right\rangle $. By repeating the Ramsey protocol many times,
allowing the interaction duration to range over several mechanical periods,
oscillations in the visibility are observed providing a signature of decoherence/recoherence.

The Ramsey protocol has the obvious advantage over the above considered
protocol (where one always remains at the $\phi_{e}=0$ degeneracy point during
$t>0$) that the decoherence/recoherence times are shorter by the factor of
$1/\zeta$. However, the disadvantage with the Ramsey protocol is that qubit
decoherence times are considerably reduced away from the degeneracy point. The
origin of the reduction in these two competing timescales is of course the
same: the mechanical oscillator and flux noise couple more strongly (i.e.,
linear coupling) to the circulating current basis states $\left\vert
\curvearrowright\right\rangle $ and $\left\vert \curvearrowleft\right\rangle $
than to the eigenstate basis states at the degeneracy point (i.e., quadratic
coupling). Depending on how the qubit decoherence rate varies with the
externally applied flux, it may be that operating a small distance from the
degeneracy point is more favorable for observing recoherence effects
\cite{Yoshihara_0606481}. However, the resulting coupled quantum dynamics is
not as simple to describe as at the special limiting bias points where the
Hamiltonian $\mathcal{H}_{1}$ [Eq. (\ref{2-level H_1})] is either
(approximately) purely diagonal or off-diagonal.

In the present paper we have considered a flux qubit in the form of an RF
SQUID, a system that is relatively simple to analyze. However, a double well
potential can be formed only when the inductance $L$ is sufficiently large and
the condition $\beta_{L}>1$ is satisfied. In this limit, the loop is
relatively large and consequently large pickup of external flux noise results
in a relatively short flux qubit decoherence time \cite{Friedman_43}. On the
other hand, this problem can be partly solved by employing the configuration
of a loop having three JJs \cite{Mooij_1036}, where a portion of the necessary
total SQUID inductance is provided by the effective inductance of the
additional JJs; the three JJ superconducting loop would likely be the
preferred choice for experimental implementation.

\section{Acknowledgements}

This work is partly supported by the US - Israel Binational Science Foundation
(BSF) and by the Israeli ministry of science. M. P. B. thanks the Aspen Center
for Physics for their hospitality and support.

\newpage
%Just because of unusual number of tables stacked at end
\bibliographystyle{apsrev}
\bibliography{acompat,Eyal_Bib}
%Produces the bibliography via BibTeX.

\end{subequations}
\end{document}